\documentclass[a4paper]{article}

\usepackage{INTERSPEECH2021}
\usepackage{float}
\usepackage{amsmath,graphicx}
\usepackage{booktabs}
\usepackage{float}
\usepackage{multirow}
\usepackage{enumitem}
\usepackage{caption}
\usepackage{subcaption}

\usepackage{amsmath}

\DeclareMathOperator*{\argmin}{arg\,min}

\title{Wav2vec-C: A Self-supervised Model for Speech Representation Learning}
\name{Author Name$^1$, Co-author Name$^2$}
\name{Samik Sadhu$^1$, Di He$^2$, Che-Wei Huang$^2$, Sri Harish Mallidi$^2$, Minhua Wu$^2$, Ariya Rastrow$^2$, \\ \textit{Andreas Stolcke$^2$, Jasha Droppo$^2$, Roland Maas$^2$}}
\address{$^1$Johns Hopkins University, USA\\  $^2$Amazon Alexa, USA}
\email{samiksadhu@jhu.edu,\{deehe,cheweh,mallidih,wuminhua,arastrow,stolcke,drojasha,rmaas\}@amazon.com}

\begin{document}

\maketitle
\begin{abstract}
Wav2vec-C introduces a novel representation learning technique combining elements from wav2vec 2.0 and VQ-VAE. Our model learns to reproduce quantized representations from partially masked speech encoding using a contrastive loss in a way similar to wav2vec 2.0. However, the quantization process is regularized by an additional consistency network that learns to reconstruct the input features to the wav2vec 2.0 network from the quantized representations in a way similar to a VQ-VAE model. The proposed self-supervised model is trained on 10k hours of unlabeled data and subsequently used as the speech encoder in a RNN-T ASR model and fine-tuned with 1k hours of labeled data. This work is one of the very few studies of self-supervised learning on speech tasks with a large volume of real far-field labeled data. The wav2vec-C encoded representations achieve, on average, twice the error reduction over baseline and a higher codebook utilization in comparison to wav2vec 2.0.
\end{abstract}

\section{Introduction}
\label{sec:intro}

Self-supervision \cite{schneider2019wav2vec,chen2020big,oord2018representation,lan2019albert} is a paradigm of machine learning (ML) that deals with unsupervised learning of structural patterns in data by exploiting contextual information.  Self-supervision has been of significant interest in the automatic speech recognition (ASR) literature primarily as a pre-training step before a fully supervised  task. In particular, it is widely used for problems with some amount of labeled data (for supervised training) and a significantly larger volume of unlabeled data (for self-supervised training). The recently proposed wav2vec 2.0 \cite{baevski2020wav2vec}  is one such self-supervised learning model that learns to predict masked out discrete speech encodings using a contextualized representation from a transformer model \cite{vaswani2017attention} . \\

In this paper, we introduce the wav2vec-C model that solves a more rigorously defined self-supervised learning problem compared to the wav2vec 2.0. In the latter, a contrastive loss defined on discretized codes drives the self-supervised learning - including the codebook in the built-in differentiable Vector Quantization module. In contrast, wav2vec-C facilitates codebook learning through an additional regularization on the discrete speech representations by reconstructing the discrete codes to the input features. Thus, wav2vec-C maintains a \textit{consistency} between the learnt representations and the input features to the network. \\

Our main contributions in this paper are
\begin{itemize}%[topsep=0pt,itemsep=-1ex,partopsep=1ex,parsep=1ex]
\item The wav2vec-C model (Section \ref{sec:model}) 
\item We use real world far-field voice query speech with varied degrees of SNR ranging between -40 to 50 dB, whereas most studies on self-supervised learning in the literature  use clean read speech \cite{chung2020generative,baevski2019vq} and some use simulated noisy speech \cite{ravanelli2020multi}.
\item Self-supervised learning has been shown to be useful for settings with little labeled data \cite{schneider2019wav2vec,ravanelli2020multi}. It has been observed that the effectiveness of self-supervision decreases as the amount of labeled data increases\cite{amazon,chung2020generative}. In this work, we explore the applicability of self-supervision with a relatively large amount of labeled data (1k hours).
\item  We also limit our model size to facilitate low-latency production level ASR models, which goes against the general trend of exceedingly large self-supervised models proposed in the literature \cite{chen2020big}.
\item We explore and compare different variants of our framework in the choice of the vector quantization framework and the effect it has on robustness and codebook utilization.
\end{itemize}

\section{The Wav2vec-C Model}
\label{sec:model}

\subsection{Summary of wav2vec 2.0}

Our model is similar to wav2vec 2.0 \cite{baevski2020wav2vec}, but differs in the way we use log short-term Fourier transform (log-STFT) features as input to our model. An \textit{encoder network} $f: \mathcal{X} \rightarrow \mathcal{Z}$ maps the input features $X=[x_1,x_2,\dots x_T]$ to a latent embedding space. These embeddings are quantized by a \textit{vector quantization} module $q: \mathcal{Z} \rightarrow \mathcal{\hat{Z}}$. The embedded vectors $Z=[z_1,z_2,\dots z_T] \in \mathcal{Z}$ are passed through a SpecAugment \cite{park2019specaugment} module that randomly masks a portion of these embeddings to generate $Z_{masked}$. These masked embeddings are fed into a \textit{context network}  $g: \mathcal{Z} \rightarrow \mathcal{C}$ that generates a set of context representation $C=[c_1,c_2, \dots c_T]$. A contrastive score between the context representations and the vector quantized embeddings $\hat{Z}=[\hat{z}_1,\hat{z}_1, \dots \hat{z}_T]$ is maximized during network training.  \\

\subsection{Wav2vec-C}

The wav2vec 2.0 model relies on a diverse set of codes correlating to the underlying speech units learned by $q$ to enable $g$ to learn good contextual representations via the contrastive loss. However, the wav2vec 2.0 problem formulation can result in several locally optimal codebooks. A few highly probable optima observed in our experiments were

\begin{itemize}
\item \textit{Voice activity detection (VAD) codebooks} - where two codes are assigned; one for speech and the other for non-speech
\item  \textit{Temporally invariant codebooks} - where the model assigns specific codes to fixed temporal locations to enable a good contrastive loss
\end{itemize}
 Our training data consists of many similar query terms occurring at fixed temporal locations which also contributed to the model assigning fixed codes at specific temporal instances via the recurrent encoder (Section \ref{sec:enc_network}) irrespective of the underlying speech sounds. Hence, the codebook learning methodology adopted for wav2vec 2.0 might not generalize well to other datasets and different model architectures, as in our case. 
\begin{figure}[H]
\centering
  \includegraphics[trim={100 200 200 100},clip,scale=0.16]{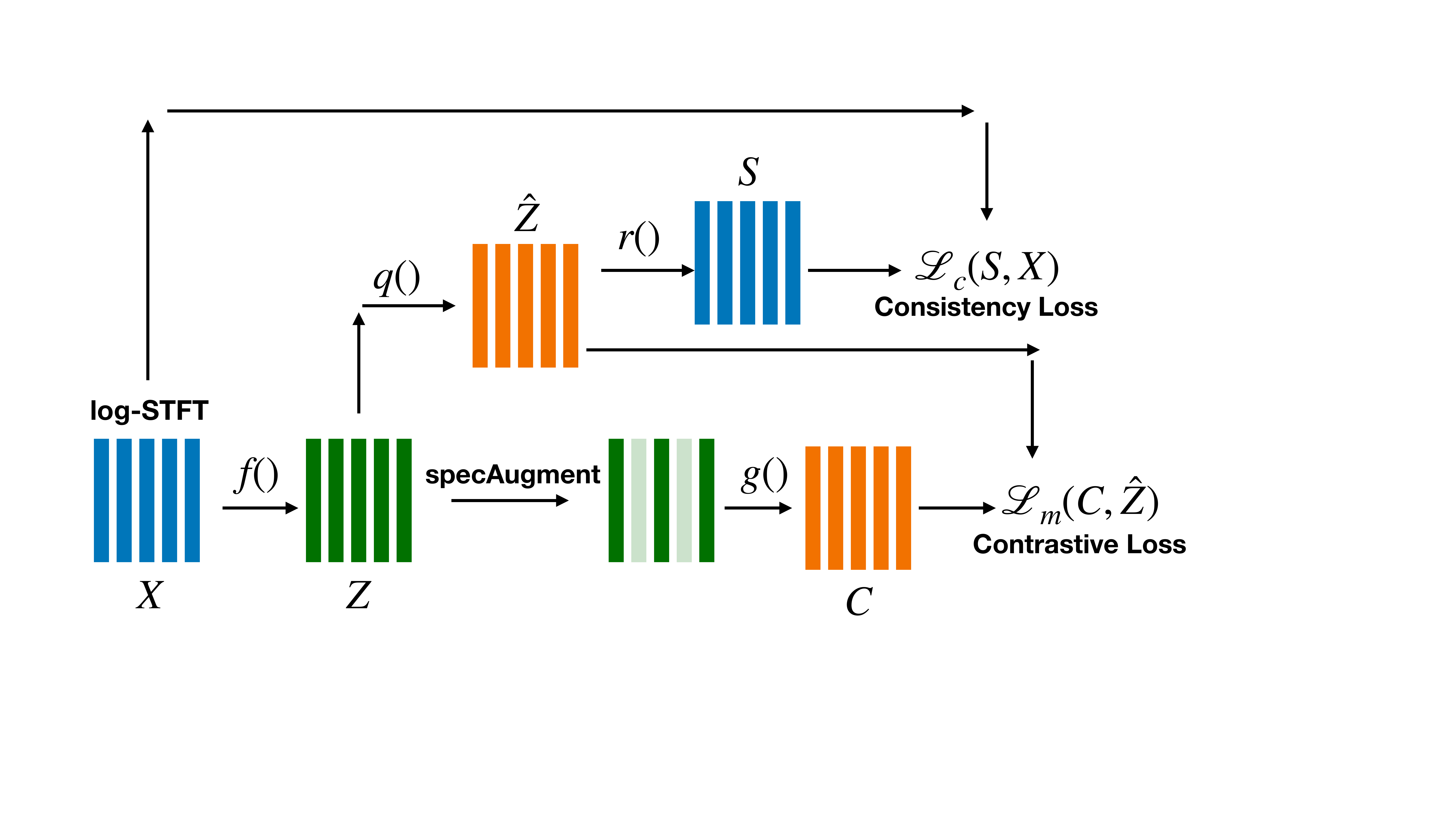} 
  \caption{Overview of the wav2vec-C self-supervised learning model}
    \label{fig:w2vc}
\end{figure}
In wav2vec-C (Figure \ref{fig:w2vc}) we enforce the codes to explicitly carry information about the input features $X$ to help mitigate the described codebook learning issues.  We define an additional  \textit{consistency network} $r: \mathcal{\hat{Z}} \rightarrow \mathcal{S}$ that reconstructs the quantized encodings $\hat{Z}=[\hat{z}_1,\hat{z}_1, \dots \hat{z}_T]$ to  consistency vectors $S=[s_1,s_1, \dots s_T]$ and minimize the normed distance between the inputs $X$ and $S$ during network training. This network allows a flow of information from the input log-STFT features back to the feature domain and enforces the latent space to preserve meaningful information that enable a low reconstruction error.  Hence, in a way, wav2vec-C can be seen as an integration of the ideas behind wav2vec 2.0 and VQ-VAE \cite{van2017neural}.

\subsection{ Encoder network ($f$)}
\label{sec:enc_network}
Our encoder network $f$ consists of three layers of long short-term memory network (LSTM) with a hidden dimension of 768. The encoder gradients are scaled by a factor $\gamma=0.1$ as in wav2vec 2.0 to help stabilize the codebook during training.

\subsection{Vector quantization ($q$)}
\label{sec:cd_loss}
We use a product quantization module \cite{baevski2019vq,jegou2010product} with $G$ codebooks $Q=[Q^{(1)},Q^{(2)}, \dots Q^{(G)}]$. Each codebook $Q^{(i)} \in \mathbb{R}^{V \times K}$ is represented by a set of $V$ codes, each of dimension $K$. The LSTM encoded representations $z \in \mathbb{R}^{768}$ are split into a set of $G$ representations $z_{split}=\{z^{(1)},z^{(2)}, \dots z^{(G)}\}$ with $z^{(i)} \in  \mathbb{R}^{768/G}, i \in \{1,2, \dots G\}$. Every $z^{(i)}$ is used to select one code $e \in \mathbb{R}^K$ from $Q^{(i)}$ to obtain a quantized representation $\hat{z}^{(i)}$. The representations $\hat{z}^{(i)}, i=\{1,2,\dots G\}$ from all the codebooks are concatenated to form the final quantized encoding $\hat{z}$. In our experiments we use $G=2$ codebooks, each with $V=320$ codes and dimension $K=384$ which is consistent with the original wav2vec 2.0 model. We use two different VQ techniques

\subsubsection{Gumbel-softmax \cite{jang2016categorical}}
Each split $z^{(i)} \in z_{split}$ is passed through a trainable linear transformation to generate logits $l^{(i)} \in \mathbb{R}^{V}$ which are passed through a Gumbel-softmax to generate a hard distribution over $V$ codes that can be used as a \textit{code selector} during the forward pass. During back propagation, we use the true gradient of the softmax distribution, thereby making the code selection process completely differentiable.

\subsubsection{K-means \cite{van2017neural}} 
During forward pass, a k-means codebook selects the code $e$ from $Q^{(i)}$ which has the closest squared distance to $z^{(i)}$ as 
\begin{equation}
\label{eq:km_forwardpass}
\hat{z}^{(i)}=\argmin_{e \in Q^{(i)}}||z^{(i)}-e||_2
\end{equation}
However, during back-propagation, a straight-through estimator \cite{bengio2013estimating} bypasses gradient computation w.r.t the quantized embedding and copies the gradients \cite{van2017neural} to the continuous embedding $z$. Since this process puts the codebook out of the training graph, there are two loss terms incorporated into training as
\begin{equation}
\label{eq:kmeans}
\mathcal{L}_{k}=||sg(z^{(i)})-\hat{z}^{(i)}||_2+\beta||z^{(i)}-sg(\hat{z}^{(i)})||_2
\end{equation}
On minimization of $\mathcal{L}_{k}$, the first term pushes the quantized representations close to the continuous encoded representation and the second term (also called \textit{commitment loss}) enforces encodings $z^{(i)}$ to commit to quantized embeddings  $\hat{z}^{(i)}$ during training. In eq. \ref{eq:kmeans}, $sg(.)$ is the stop gradient operator \cite{baevski2019vq} and $\beta=0.25$ as is the optimal value reported in \cite{van2017neural}. 

\subsection{Masking}
We use a SpecAugment \cite{park2019specaugment} module to mask out portions of the continuous encodings $Z=[z_1,z_2,\dots,z_T]$ before feeding them to the context network.  We use five masks for every utterance. Each mask has maximum width of 16\% of the utterance length. On average 40\% of the encoded frames are masked. 

\subsection{Context network ($g$)}
The context network consists of five transformer layers, with model dimension 1024 and inner feed-forward dimension of 4096 with 16 attention heads. We use sinusoidal positional embedding for the transformer layers. A contrastive score between the context representations $C=[c_1,c_2,\dots c_T]$ and the quantized encodings $\hat{Z}=[\hat{z}_1,\hat{z}_2,\dots \hat{z}_T]$  is computed as

\begin{equation}
\label{eq:contrastive_loss}
\mathcal{L}_{m}=-\log\frac{\exp(d(c_t,\hat{z}_t))/\kappa}{\sum_{z \in \Theta} \exp(d(c_t,z))/\kappa }
\end{equation}
where $t \in \{1,2,\dots T\}$,  $\Theta$ is a set consisting of $\hat{z}_t$ and a selection of $N$ negative samples, $\kappa$ is the temperature variable and $d$ calculates the cosine similarity $d(x,y)=\frac{x^Ty}{||x||||y||}$. In our experiments, we uniformly sample $N=50$ negative samples from the encodings $\hat{Z}$ of the utterance and $\kappa$ is updated as proposed in \cite{baevski2020wav2vec}.

\subsection{Consistency network ($r$)}
The consistency network $r$ consists of a 3-layer LSTM that maps the quantized embedding $\hat{Z}=[\hat{z}_1,\hat{z}_2,\dots \hat{z}_T]$ to the consistency vectors $S=[s_1,s_1, \dots s_T]$. We minimize the $L_2$ normed distance between $S$ and $X$ as 

\begin{equation}
\mathcal{L}_c=||x_t-s_t||_2
\end{equation}
\vspace{-20pt}
\subsection{Loss}
During training, we minimize the primary contrastive loss together with a codebook loss component and the consistency loss as
 \begin{equation}
\label{eq:loss}
\mathcal{L}=\mathcal{L}_{m}+ \mathcal{L}_{cb}+\gamma \mathcal{L}_c
\end{equation}
The codebook loss $\mathcal{L}_{cb}$ (section \ref{sec:cd_loss})  takes a different form according to the type of VQ used. Wav2vec 2.0 and wav2vec-C are generalized by the parameter $\gamma$, where a value $\gamma=0$ results in the wav2vec 2.0 model as the consistency loss is ignored for model training, while $\gamma=1$ leads to our wav2vec-C model in full effect. \\

For a Gumbel-softmax VQ module, the codebook loss is given by $\mathcal{L}_{cb}=\alpha\mathcal{L}_d$, where $ \mathcal{L}_{d}$ is a diversity loss on the Gumbel-softmax distribution given by 
\begin{equation}
\mathcal{L}_{d}=\frac{GV-\sum_{g=1}^G\exp(-\sum_{v=1}^Vp_{g,v}\log p_{g,v})}{GV}
\end{equation}
where $p_{g,v}$ is the probability assignment by the $g^{th}$ codebook on the $v^{th}$ code. The weight $\alpha$ on the diversity loss determines the relative importance of the component and is instrumental in avoiding the codebook collapse that is commonly observed in VQ problems \cite{chorowski2019unsupervised,van2017neural}. In our experiments, we found $\alpha=1.5$ to be suitable to avoid catastrophic codebook collapse issues.  For k-means VQ, the codebook loss is simply equal to the k-means loss, i.e., $\mathcal{L}_{cb}=\mathcal{L}_k$

\section{Experimental Setup}

\subsection{Data sets}
The goal of this study is to evaluate the effectiveness of self-supervised pre-training for real world applications.  Hence,  instead of using publicly available clean read speech we use in-house training and evaluation data consisting of \textit{real-world far-field} English voice command and voice query speech collected from home environments similar to \cite{van2019dipco} with varying degrees of SNR in the range -40 to 50 dB.  
\subsubsection{Training data}
We use 10k hours of unlabeled and 1k hours of transcribed de-identified English language training data collected from native and non-native English speakers. To our knowledge, this work is one of the first few instances where a large proportion of labeled data is used alongside self-supervised pre-training for ASR tasks, especially realistic speech queries instead of clean read speech data. 
\subsubsection{Test data}
We test our ASR models on four different test sets summarized in Table  \ref{tab:test_sets}

\setlength{\tabcolsep}{1pt}
\begin{table}[H]
\caption{Different test sets for RNN-T model evaluation}
\begin{tabular}{@{}llc@{}}
\toprule
Test set           & Details                              & \multicolumn{1}{l}{\# Utterances} \\ \midrule
clean(\tiny{$SNR_{20}$}\normalsize)   & Average SNR $\approx$ 20 dB                & 118.0k                            \\
clean(\tiny{$SNR_{16}$}\normalsize)  & Average SNR $\approx$ 16   dB             & 43.2k                             \\
noisy(\tiny{$N_1$}\normalsize)     & background multimedia speech         & 31.0k                             \\
noisy(\tiny{$N_2$}\normalsize)       & background speech, multiple speakers & 5.8k                              \\ \bottomrule
\end{tabular}

\label{tab:test_sets}
\end{table}
\subsection{Recurrent Neural Network Transducer (RNN-T) Model}
RNN-T \cite{graves2012sequence,guo2020efficient,li2019improving} ASR models are widely used for deployable end-to-end speech recognition systems because of their fast online streaming capability. We use the pre-trained wav2vec-C and wav2vec 2.0 models to initialize the speech encoder for a RNN-T ASR model.

\subsubsection{Pre-trained RNN-T}
After training the self-supervised model on unlabeled data, we use the output of the context network $g$ as speech representations. Thus, the RNN-T \textit{speech encoder} consists of three LSTM layers followed by five layers of transformer extracted from the self-supervised model with the masking module eliminated. We use two LSTM layers with 1024 hidden units as the RNN-T \textit{prediction network} and a simple single layer feedforward \textit{joint network}. The pre-trained speech encoder is also fine-tuned during RNN-T training. We use a total of 4000 sub-word tokens together with a blank token to generate the targets for RNN-T training. The RNN-T network is also regularized with SpecAugment on the input features with 10\% of the temporal frames and 30\% of the frequency bins randomly masked with noise. 25\% dropout is applied on the transformer weights.

\subsubsection{Baseline RNN-T}
Our baseline model consists of an RNN-T with the same architecture as the pre-trained model but without pre-training the speech encoder. 

\subsection{Training details}
We train 4 different self-supervised models
\begin{enumerate}%[topsep=0pt,itemsep=-1ex,partopsep=1ex,parsep=1ex]
\item wav2vec 2.0 (GS): $\gamma=0$, Gumbel-softmax codebook
\item  wav2vec 2.0 (KM): $\gamma=0$, k-means codebook
\item  wav2vec-C (GS): $\gamma=1$, Gumbel-softmax codebook
\item  wav2vec-C (KM): $\gamma=1$, k-means codebook
\end{enumerate}
Subsequently, we train RNN-T models with the speech encoder replaced by the self-supervised models. \\

Our models are trained using Tensorflow 2.0. The self-supervised models are trained for 100k steps with 30 minutes of speech per step. We use an Adam optimizer \cite{kingma2014adam}, where the learning rate is warmed up from $1\times 10^{-7}$ and held at $5\times 10^{-6}$ after 3k steps. The RNN-T models are trained for 60k steps, with an average of 1 hours of speech per step. The learning rate is warmed up from $1\times 10^{-7}$ and held at $5\times 10^{-4}$ after 3k steps.

\section{Results}
We compare the word error rate reduction \textit{relative to the baseline model} (rWERR) for the different pre-trained RNN-T models evaluated on the four test sets in Table \ref{tab:main_table}. The baseline ASR model has $<10\%$ absolute word error rate. To smooth out error fluctuations, we report the mean rWERR computed after 50k, 55k and 60k RNN-T training steps. The average rWERR in the last column is the rWERR for each test set \textit{weighted} by the number of utterances in that test set.

Our implementation of the wav2vec 2.0 pre-trained RNN-T model does not show noticeable performance improvement over baseline for the clean test sets. Whereas, for the noisy test sets, some gains can be observed - with wav2vec 2.0 (KM) performing better, on average, compared to wav2vec 2.0 (GS). This trend is comparable to the results reported in \cite{baevski2020wav2vec}, where pre-training is shown to be most beneficial for the challenging \textit{test\_other} test set of Librispeech \cite{panayotov2015librispeech}. However, while drawing this comparison we should keep in mind the major differences between the best performing wav2vec 2.0 models in \cite{baevski2020wav2vec} and our implementation, namely

\begin{enumerate}[topsep=0pt,itemsep=-1ex,partopsep=1ex,parsep=1ex]
	\item We use a much smaller context network (5 layers) compared to the original (24 layers)
	\item We use a 3-layer LSTM as encoder with log-STFT input features
\end{enumerate}

\setlength{\tabcolsep}{3pt}
\renewcommand{\arraystretch}{1.2}
\begin{table}[H]
\centering
\caption{Comparison of rWERR of different pre-trained RNN-T models. \\
(*The errors on each test set weighted by the number of utterances in each test set)}
\begin{tabular}{@{}lccccc@{}}
\toprule
\multirow{2}{*}{RNN-T encoder} & \multicolumn{5}{c}{rWERR on different test sets} \\ \cmidrule(l){2-6} 
                               & $SNR_{20}$   & $SNR_{16}$   & $N_1$   & $N_2$  & Average$^{*}$  \\ \midrule
\textbf{wav2vec 2.0 (KM)}       & 0          & 0.6        & 0.7     & 3.2    & 0.7  \\
\textbf{wav2vec 2.0 (GS)}      & 0          & 0.6        & 0.7     & 2.7    & 0.3  \\
\textbf{wav2vec-C (KM)}         & 0.8        & 0.6        & 0.7     & 2.7    & 0.8  \\
\textbf{wav2vec-C (GS)}         & 1.6        & 1.2        & 0.7     & 1.6    & 1.4  \\ \bottomrule
\end{tabular}

\label{tab:main_table}
\end{table}

The wav2vec-C encoded RNN-T models, on the other hand, show a positive rWERR for both $SNR_{20}$ as well as $SNR_{16}$ clean test sets. In particular, wav2vec-C (GS) gains 1.6\% rWERR on $SNR_{20}$ and 1.2\%  rWERR on $SNR_{16}$. However, there is a reduction in performance (in comparison to wav2vec 2.0) for the noisy test sets. This suggests that the reconstruction idea adopted for wav2vec-C leads to an overall better performance of the pre-trained RNN-T model, however with a slight loss in robustness.

\subsection{Discussions on codebook utilization}
Our codebooks have a maximum capacity of $320 \times 320=102.4$k codes with wav2vec-C (GS) utilizing the full 100\% of the codebook (see Table \ref{tab:2}).  Hence, the consistency loss together with the weight $\alpha$ on the diversity loss enforces the model to pick a variety of codes to minimize the reconstruction loss.

\begin{table}[H]
\centering
\caption{Codebook utilization of the self-supervised models}
\begin{tabular}{@{}lc@{}}
\toprule
self-supervised model     & \begin{tabular}[c]{@{}c@{}}codebook \\ utilization(\%)\end{tabular} \\ \midrule
\textbf{wav2vec 2.0 (KM)} &  $<1$                                                        \\
\textbf{wav2vec 2.0 (GS)} & $15$                                                         \\
\textbf{wav2vec-C (KM)}   & $<1$                                                                 \\
\textbf{wav2vec-C (GS)}   & $100$                                                        \\ \bottomrule

\end{tabular}

\label{tab:2}
\end{table}

\begin{figure}[H]
    \centering
    \begin{subfigure}[t]{0.5\textwidth}
        \centering
        \includegraphics[trim=0 0 0 0, clip,scale=0.12]{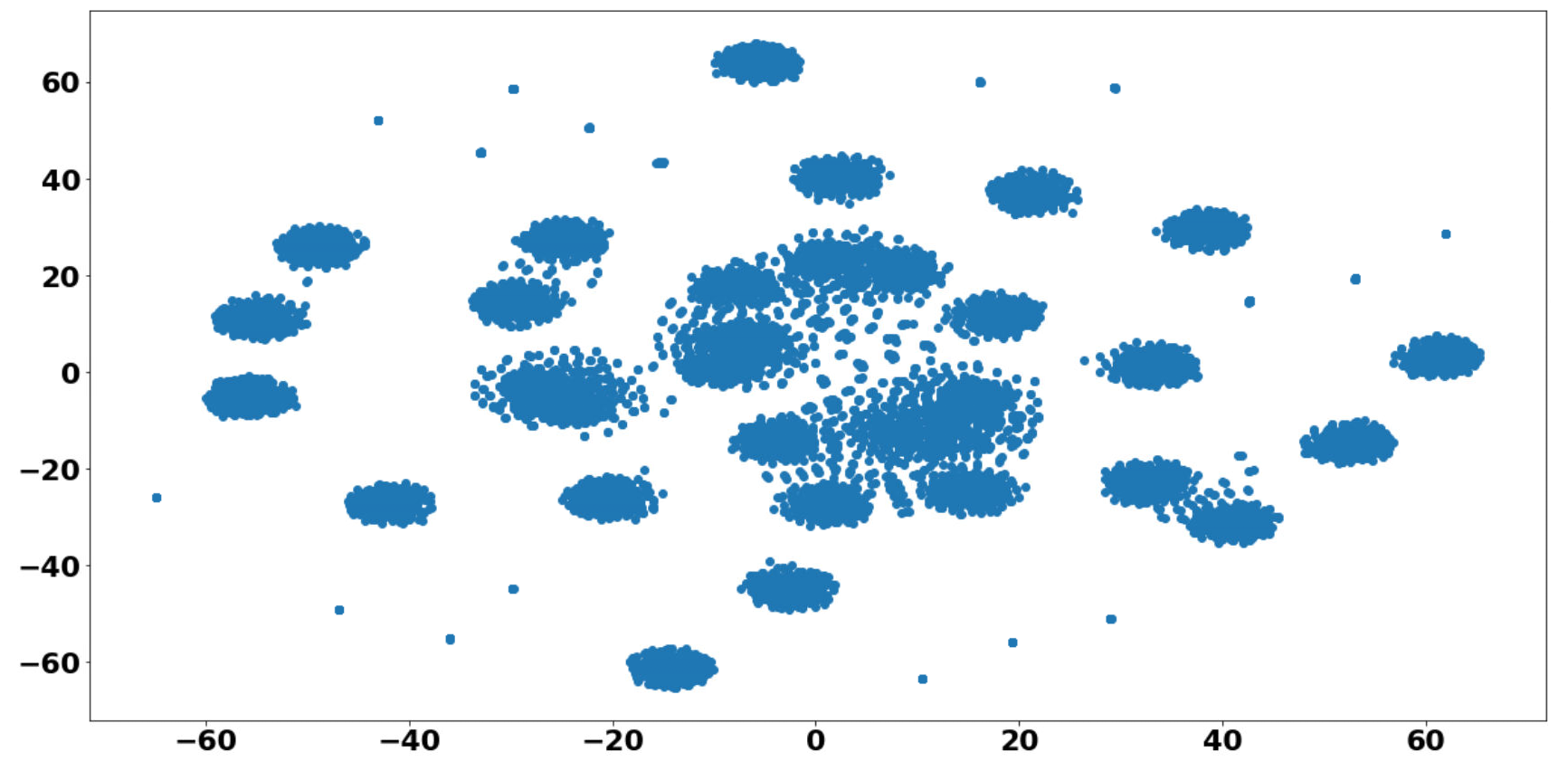} 
        \caption{wav2vec 2.0 (GS)}
         \label{fig:t-SNEa}
    \end{subfigure}%
    \\
    \begin{subfigure}[t]{0.5\textwidth}
        \centering
        %\hspace{-20pt}
        \includegraphics[trim=0 0 0 0, clip,scale=0.12]{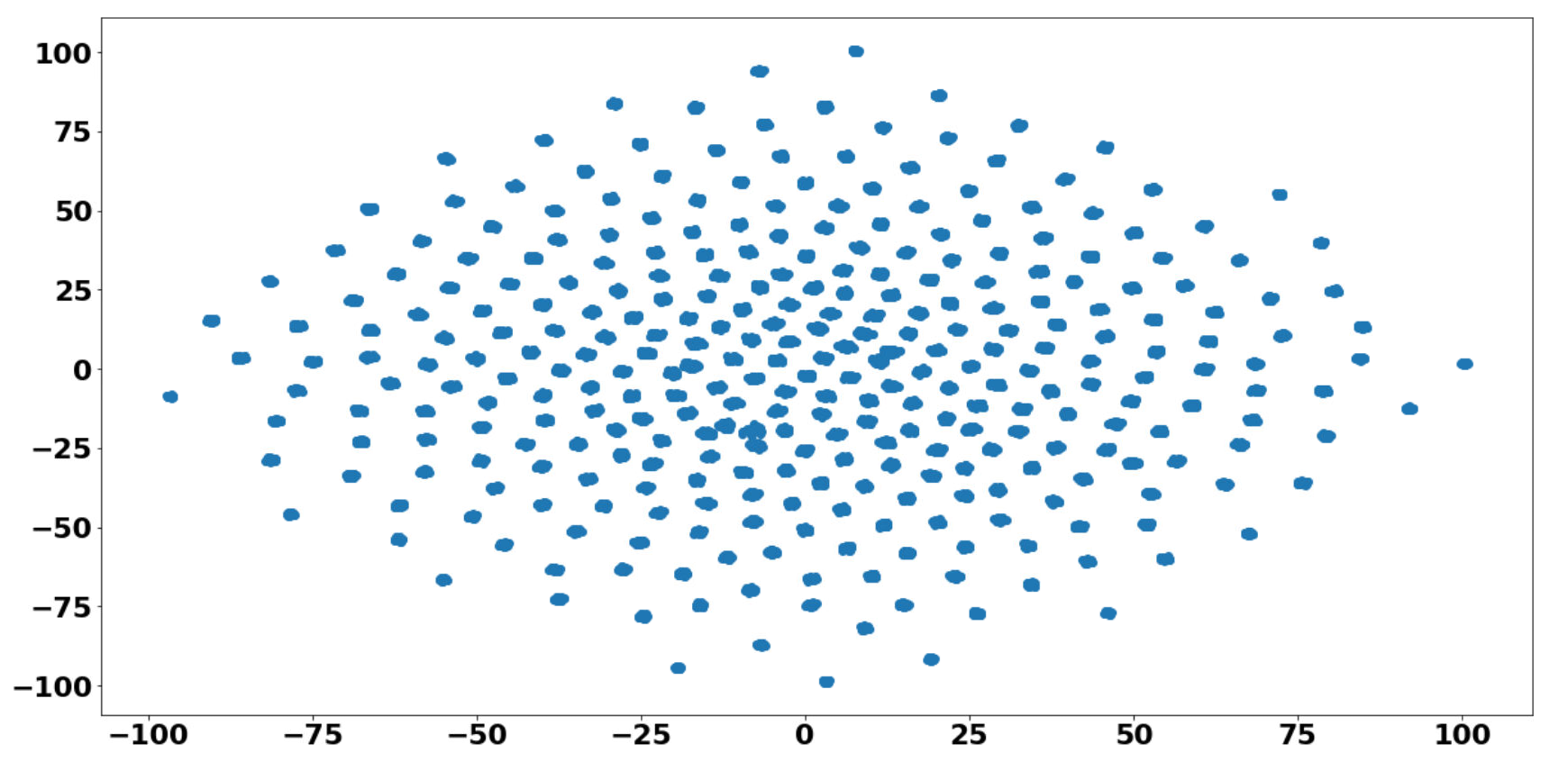} 
        \caption{wav2vec-C (GS)}
        \label{fig:t-SNEb}
    \end{subfigure}
    \caption{t-SNE clustering of the wav2vec 2.0 (GS) and wav2vec-C (GS) codebooks}
    \label{fig:t-SNE}
\end{figure}

A t-SNE plot of the 102.4k codes in the 100\% utilized codebook of the wav2vec-C (GS) model can be seen in Figure \ref{fig:t-SNEb} showing the clusters formed by the codes over the course of training. On the other hand, the 102.4k codes learnt by wav2vec 2.0 (GS), as shown in Figure \ref{fig:t-SNEa}, form a smaller number of clusters with significant inter-cluster overlap possibly due to the under-utilized codebook.

The k-means codebook uses only a small fraction of the codes but is more robust compared to Gumbel-softmax models for noisy test sets, in particular the $N_2$ noisy test set. For example, a comparison of the ASR performances of wav2vec-C (GS) and wav2vec-C (KM) would show that wav2vec-C (GS) gives a better rWERR for clean test sets in comparison to noisy test sets, whereas wav2vec-C (KM) shows the opposite characteristics. This observation highlights the importance of codebook diversity for different application domains. For example, a small codebook diversity is not necessarily a bad design choice if robustness is of importance during model evaluation.

\section{Conclusions}
In this paper we propose wav2vec-C, a new self-supervised learning model which is based on an amalgamation of the ideas from wav2vec 2.0 and VQ-VAE with the goal of solving the codebook utilization difficulties observed for wav2vec 2.0.  We used real-world far-field noisy data for self-supervised learning and 1k hours of data for supervised ASR training. The proposed self-supervised model after RNN-T fine-tuning achieved, on average, a 1.4\% relative WER reduction over baseline compared to a 0.7\% reduction from wav2vec 2.0. Furthermore, we also observed that ASR robustness is correlated with codebook diversity, validating our motivation for the wav2vec-C architecture
\newpage

\bibliographystyle{IEEEtran}

\bibliography{mybib}

% \begin{thebibliography}{9}
% \bibitem[1]{Davis80-COP}
%   S.\ B.\ Davis and P.\ Mermelstein,
%   ``Comparison of parametric representation for monosyllabic word recognition in continuously spoken sentences,''
%   \textit{IEEE Transactions on Acoustics, Speech and Signal Processing}, vol.~28, no.~4, pp.~357--366, 1980.
% \bibitem[2]{Rabiner89-ATO}
%   L.\ R.\ Rabiner,
%   ``A tutorial on hidden Markov models and selected applications in speech recognition,''
%   \textit{Proceedings of the IEEE}, vol.~77, no.~2, pp.~257-286, 1989.
% \bibitem[3]{Hastie09-TEO}
%   T.\ Hastie, R.\ Tibshirani, and J.\ Friedman,
%   \textit{The Elements of Statistical Learning -- Data Mining, Inference, and Prediction}.
%   New York: Springer, 2009.
% \bibitem[4]{YourName17-XXX}
%   F.\ Lastname1, F.\ Lastname2, and F.\ Lastname3,
%   ``Title of your INTERSPEECH 2021 publication,''
%   in \textit{Interspeech 2021 -- 20\textsuperscript{th} Annual Conference of the International Speech Communication Association, September 15-19, Graz, Austria, Proceedings, Proceedings}, 2020, pp.~100--104.
% \end{thebibliography}

\end{document}